\documentclass[11pt]{article}
\usepackage{amssymb}
\usepackage{amsmath}
\usepackage{graphics}
\usepackage{epsfig}
\usepackage{a4wide}
\usepackage{cite}
\usepackage{multirow}
\usepackage{color}
\usepackage{mathrsfs}
\usepackage{slashed}

\DeclareGraphicsExtensions{.eps}
\begin{document}

\title{$Z_H \rightarrow H^0 \gamma$ decay within the littlest Higgs model at $\mathcal{O}(\alpha_{{\rm ew}}^{3}\alpha_s)$ accuracy}
\author{
Ming-Ming Long,$^{1,2}$ Shao-Ming Wang,$^{3}$ Qiang Yang,$^{1,2}$ Ren-You Zhang,$^{1,2}$\footnote{Corresponding author: zhangry@ustc.edu.cn} \\
Wen-Gan Ma,$^{1,2}$ and Jian-Wen Zhu$^{1,2}$ \\ \\
{\small $^1$ State Key Laboratory of Particle Detection and Electronics,} \\
{\small University of Science and Technology of China, Hefei 230026, Anhui, People's Republic of China} \\
{\small $^2$ Department of Modern Physics, University of Science and Technology of China,}  \\
{\small Hefei 230026, Anhui, People's Republic of China} \\
{\small $^3$ Department of Physics, Chongqing University, Chongqing 401331, People's Republic of China}  \\
}

\date{}
\maketitle
\vskip 10mm

\begin{abstract}
We study the rare decay $Z_H \rightarrow H^0 \gamma$ at $\mathcal{O}(\alpha_{{\rm ew}}^{3}\alpha_s)$ accuracy including two-loop QCD corrections in the context of the littlest Higgs model (LHM) without $T$-parity. We revisit analytically and numerically the leading-order (LO) contributions of the one-loop diagrams induced by massive fermions, scalars and charged gauge bosons in the LHM, and further study the NLO QCD correction to this decay process. We perform the numerical calculation by taking the LHM input parameters $f=3,~4~\text{TeV}$ and $0.1< c <0.6$, and discuss the numerical results of the decay width up to the QCD NLO within the recent experimentally constrained LHM parameter space region. Our results show that the two-loop QCD correction always reduces the LO decay width and the top-induced QCD correction is the dominant contribution at the QCD NLO. For $f=4~\text{TeV}$ and $c=0.3$, the NLO QCD corrected decay width reaches $75.099~\text{keV}$ and the NLO QCD relative correction is about $-11.0\%$.
\end{abstract}


\vfill \eject
\baselineskip=0.32in
\makeatletter      
\@addtoreset{equation}{section}
\makeatother       
\vskip 5mm
\renewcommand{\theequation}{\arabic{section}.\arabic{equation}}
\renewcommand{\thesection}{\Roman{section}.}
\newcommand{\nb}{\nonumber}
\vskip 5mm
\section{Introduction}
\par
Although the standard model (SM) \cite{s1,s2} has got a remarkable success in describing high-energy phenomena at the energy scale up to $10^2~ {\rm GeV}$, the mechanism of electroweak symmetry breaking (EWSB) remains the most prominent mystery, and the Higgs boson mass suffers from large radiative corrections in the SM. Alternatively, in the little Higgs (LH) models \cite{LHMs,Low} based on dimensional deconstruction \cite{Decon}, the quadratic divergence induced at the one-loop level by the SM gauge bosons is cancelled by the heavy gauge boson one-loops. Therefore, there arouse more and more interests on the LH models as they offer an alternative approach to solve the hierarchy problem, and the LH models were proposed as one kind of models of EWSB without fine-tuning in which the Higgs boson is naturally light as a result of non-linearly realized symmetry \cite{LHMs,Low,LH4,LH6,LH7}.

\par
The most economical model of them is the littlest Higgs model (LHM), which is based on an $SU(5)/SO(5)$ nonlinear sigma model \cite{Low}. In the LHM without $T$-parity, in addition to the SM particles, a set of new heavy gauge bosons ($A_H,~Z_H,~W_H$) and an exotic heavy vector-like quark ($T$) are introduced which just cancel the quadratic divergences induced by the SM gauge boson loops and the top quark loop, respectively. The key feature of this model is that the Higgs boson is a pseudo-Goldstone boson of a global symmetry, which is spontaneously broken at some higher scale $f$, and thus the Higgs boson is naturally light. On the other hand, there are also several other models that predict the existence of a neutral massive gauge boson, identified as $Z^\prime$ gauge boson, such as the 331 model \cite{331} and the grand unified models \cite{GUT}. This type of particles are under exhaustive search at the LHC \cite{Zpatlas-mass,Zpcms-mass,Atlas-1}, where the ATLAS and CMS collaborations have imposed experimental bounds over the mass of a new particle related to the $Z^\prime$ gauge boson.

\par
It is well known that the parameters of the LHM without $T$-parity are very constrained by the electroweak precision observables\cite{Csaki,Hewett}, such as $Z$-boson mass and partial widths for $Z$ decaying into lepton or light hadron pairs, since the new heavy particles predicted by the LHM can contribute to those SM processes at the tree level via $s$-channel exchange. Thus, we may expect that the virtual effects on all other SM processes induced by the exchange of new heavy particles are also negligible after considering the stringent constraints on the parameter space. However, the characteristic signal processes of the LHM, such as the productions of new heavy gauge bosons and their decays, are not very severely restricted by those constraints from the electroweak precision observables. Thus, it is still worthwhile to study them in considerable detail within the framework of the LHM without $T$-parity.

\par
The $Z_H \rightarrow H^0\gamma$ decay process can be used to identify the production of the $Z_H$ gauge boson at high energy colliders, since recent measurements on the Higgs boson discovery channels and electroweak precision observables have provided severe constraints on its parameter space \cite{Reuter}. Another advantage in probing the $Z_H \rightarrow H^0\gamma$ decay channel is due to the fact the SM background is naturally suppressed \cite{Abbasabadi:1997zr,Passarino:2013nka}. Therefore, $H^0 \gamma$ associated production at high energy colliders opens a new window to test the gauge sector of the SM and Higgs physics \cite{Toscano,Martinez,Aad:2010sp,Aad:2013zba}. Previous study on the $Z^\prime \rightarrow H^0\gamma$ decay has been performed in the context of left-right symmetric models \cite{Toscano}, where the branching ratio is estimated \cite{Aranda}. In this paper, we investigate the QCD two-loop correction to the $Z_H\rightarrow H^0\gamma$ decay and provide the decay width up to the $\cal{O}$$(\alpha_{{\rm ew}}^{3}\alpha_s)$ in the LHM.

\par
The rest of this paper is organized as follows. In Sec.II we briefly review the LHM. In Sec.III we present the analytical calculation at the LO and QCD NLO for the $Z_H \rightarrow H^0 \gamma$ decay in the LHM without $T$-parity. The numerical results and discussion are provided in Sec.IV. Finally, we give a short summary in Sec.V.

\section{Related theory of LHM}
\par
The LHM is based on an $SU(5)/SO(5)$ nonlinear sigma model. The nonlinear sigma model $SU(5)$ symmetric tensor field $\Sigma$ is parameterized as
\begin{eqnarray}
\Sigma(x) = e^{i \Pi(x)/f} \Sigma_0 e^{i \Pi(x)^{T}/f},
\end{eqnarray}
where the vacuum expectation value (VEV) of $\Sigma(x)$ is given by \cite{Low,LH8}
\begin{eqnarray}
\Sigma_0 =
\langle \Sigma \rangle
=
\left(
\begin{array}{ccc}
&   & {\bf{1}}_{2 \times 2} \\
& 1 & \\
{\bf{1}}_{2 \times 2} & &
\end{array}
\right).
\end{eqnarray}
At the energy scale $f \sim \mathcal{O}({\rm TeV})$, the $SU(5)$ global symmetry breaks down to its $SO(5)$ subgroup, and the $[SU(2) \otimes U(1)]^2$ gauge subgroup of $SU(5)$ simultaneously breaks down to its diagonal subgroup $SU(2)_L \otimes U(1)_Y$, which is identified as the SM electroweak gauge group. The $SU(5)/SO(5)$ symmetry breaking leads to 14 massless Nambu-Goldstone bosons. The Goldstone boson matrix is written as $\Pi(x) = \pi^a(x) X^a$. $X^a$ are the broken generators of $SU(5)$ which satisfy the relation
\begin{eqnarray}
X^a\Sigma_0-\Sigma_0 X^{aT}=0.
\end{eqnarray}
Then the Goldstone boson matrix $\Pi(x)$ can be expressed as
\begin{eqnarray}
\Pi=\left(
\begin{array}{ccc}
& h^\dagger/\sqrt 2 & \phi^\dagger \\
h/\sqrt 2 &  & h^*/\sqrt 2\\
\phi  &  h^T/\sqrt 2 &
\end{array}
\right),
\end{eqnarray}
where $h$ and $\phi$ are the SM $SU(2)_L$ doublet and triplet, respectively, and can be expressed as
\begin{eqnarray}
h = \left(
    \begin{array}{cc}
    h^{+} ~~ h^0
    \end{array}
    \right), ~~~~~~
\phi = \left(
       \begin{array}{cc}
       \phi^{++} & \phi^{+}/\sqrt{2} \\
       \phi^{+}/\sqrt{2} & \phi^0
       \end{array}
       \right).
\end{eqnarray}
The leading order dimension-two term for the scalar field $\Sigma(x)$ in the LHM is given by
\begin{eqnarray}
{\cal L} = \frac{1}{2} \frac{f^2}{4} {\rm Tr} |{\cal D}_{\mu}\Sigma|^2.
\end{eqnarray}
${\cal D}_{\mu}$ is the covariant derivative for gauge group $[SU(2) \otimes U(1)]^2 = [SU(2)_1 \otimes U(1)_1] \otimes [SU(2)_2 \otimes U(1)_2]$, and we have
\begin{eqnarray}
{\cal D}_\mu \Sigma =
\partial_\mu \Sigma - i \sum_{j=1}^2
\left[
g_j \sum_{a=1}^3 W_{\mu j}^a (Q_j^a \Sigma + \Sigma Q_j^{aT})
+
g_j^{\prime}B_{\mu j}(Y_j \Sigma + \Sigma Y_j^T)
\right],
\end{eqnarray}
where $W_{\mu j}^a$ and $B_{\mu j}$ are the $SU(2)_j$ and $U(1)_j$ gauge fields, respectively. The generators of the $SU(2)_j$ and $U(1)_j$ gauge groups are written as
\begin{eqnarray}
&&
Q_1^a = \left(
        \begin{array}{cc}
        \dfrac{\sigma^a}{2} & \\
        & {\bf{0}}_{3 \times 3}
        \end{array}
        \right),
~~~~~~~\,
Y_1 = {\rm diag}\{-3,~ -3,~ 2,~ 2,~ 2\}/10,
\nonumber \\
&&
Q_2^a = \left(
        \begin{array}{cc}
        {\bf{0}}_{3 \times 3} & \\
        & -\dfrac{\sigma^{a \ast}}{2}
        \end{array}
        \right),
~~~~
Y_2 = {\rm diag}\{-2,~ -2,~ -2,~ 3,~ 3\}/10,
\end{eqnarray}
where $\sigma^a~ (a = 1, 2, 3)$ are the Pauli matrices. As we know, in the LHM there is no Higgs potential at tree-level. Instead, the Higgs potential is generated at one-loop and higher orders due to the interactions with gauge bosons and fermions. The Higgs potential (Coleman-Weinberg potential) up to the operators of dimension four can be expressed as \cite{LH8,LH9}
\begin{eqnarray}
\label{win} V
&=&
\lambda_{\phi^2} f^2 {\rm Tr}( \phi^{\dag} \phi) + i \lambda_{h \phi h} f (h \phi^{\dag} h^T - h^{\ast} \phi h^{\dag}) - \mu^2 h h^{\dag} + \lambda_{h^4} (h h^{\dag})^2 \nonumber \\
&&
+\, \lambda_{h \phi \phi h} h \phi^{\dag} \phi h^{\dag} + \lambda_{h^2 \phi^2} h h^{\dag} {\rm Tr}(\phi^{\dag} \phi) + \lambda_{\phi^2 \phi^2} \left( {\rm Tr}(\phi^{\dag} \phi) \right)^2 \nonumber \\
&&
+\, \lambda_{\phi^4} {\rm Tr}(\phi^{\dag} \phi \phi^{\dag} \phi).
\end{eqnarray}
By minimizing the Coleman-Weinberg potential, we obtain $\langle h^0 \rangle =v/\sqrt 2$ and $\langle i \phi^0 \rangle=v^{\prime}$, which give rise to the EWSB.
After the EWSB, the gauge sector acquires additional mass and mixing term due to the VEVs of $h$ and $\phi$. By diagonalizing the quadratic term of the gauge sector, we may get the mass eigenstates $A_L$, $Z_L$, $W_L$, $A_H$, $Z_H$ and $W_H$, and their masses.

\par
To avoid large quadratic divergence in the Higgs boson mass due to the top Yukawa interaction, we introduce a pair of new fermions $\tilde{t}$ and $\tilde{t}^{\prime}$ \cite{LH8} and a set of new interactions. The scalar couplings to the top quark can be taken from the following Lagrangian \cite{LH8}:
\begin{eqnarray}
\label{L1}{\cal L}_Y
=
\frac{1}{2}\lambda_1 f \epsilon_{ijk}\epsilon_{xy}\chi_i \Sigma_{jx}\Sigma_{ky}u^{\prime c}_3+ \lambda_2f \tilde{t}\tilde{t}^{\prime c}+h.c.,
\end{eqnarray}
where $\chi=(b_3,\, t_3,\, \tilde{t})$, $\epsilon_{ijk}$ and $\epsilon_{xy}$ are antisymmetric tensors with $i,\, j,\, k \in \{1,\, 2,\, 3\}$ and $x,\, y \in \{4,\, 5\}$, and the coupling constants $\lambda_1$ and $\lambda_2$ are supposed to be of the order of unity. After expanding the above Lagrangian and performing field redefinition\cite{LH8,Buras}, we get the SM top quark $t$ and a new heavy vector-like quark $T$. The masses of the two mass eigenstates are given by
\begin{eqnarray}
\label{eq:MT}
&& m_t = c_{\lambda}^2 \lambda_2 v \left\{ 1 + \frac{v^2}{f^2} \left[ -\frac{1}{3} + \frac{x}{4}
         + \frac{1}{2} c_{\lambda}^2 \left(1 - c_{\lambda}^2 \right) \right] \right\},
 \\
\label{eq:MTH}
&& m_T = \frac{\lambda_2 f}{\sqrt{1-c_{\lambda}^2}} \left[ 1 - \frac{v^2}{f^2}\frac{1}{2}c_{\lambda}^2(1-c_{\lambda}^2) \right],
\end{eqnarray}
where $c_{\lambda} = \dfrac{\lambda_1}{\sqrt{\lambda_1^2+\lambda_2^2}}$ and $x = 4f\dfrac{v^{\prime}}{v^2}$. Considering the EWSB, we may obtain the masses of the new heavy gauge bosons and scalars as \cite{Buras}
\begin{eqnarray}
&&
m_{W_H^{\pm}}^2 = m_W^2\left( \dfrac{f^2}{s^2c^2v^2}-1\right), ~~~~~~~~~~
m_{Z_H}^2 = m_W^2\left( \dfrac{f^2}{s^2c^2v^2}-1\right),
\label{MZH-mass} \\
&&
m_{A_H}^2 = m_Z^2 s_W^2\left( \dfrac{f^2}{5s^{\prime\,2}c^{\prime\,2}v^2}-1\right), ~~~~
m^2_\Phi = \frac{2m_H^2 f^2}{v^2}\dfrac{1}{\left(1- x^2 \right)}.
\end{eqnarray}

\section{Calculation strategy}
\subsection{General setup}
\par
We employ the modified FeynArts-3.9 package \cite{FeynArts} to generate all the one- and two-loop Feynman diagrams and their corresponding amplitudes. The reduction of output amplitudes is accomplished by the FeynCalc-9.0 package \cite{FeynCalc-1,FeynCalc-2}. In our one- and two-loop amplitude calculation, we apply the FIRE \cite{Alexander-7} and Reduze2 \cite{reduze2} packages, in which the integration-by-parts (IBP) identities and Lorentz invariance (LI) identities are adopted, to perform the loop reduction and express the amplitude in terms of a certain number of independent master integrals (MIs) depending on the loop order. A scalar multi-loop integral in $d=4-2\epsilon$ dimensions is defined as
\begin{eqnarray}
G(a_1, ..., a_n) = \int \prod_{i=1}^{L} \frac{d^d{l_i}}{(2\pi)^d} \frac{1}{\prod_j^n D_j^{a_j}},
\end{eqnarray}
where $L$ is the number of loops, $l_i$ is the $i$-th loop momentum, $n$ is the number of independent propagators, and $a_j \in \mathbb{Z}$. The $j$-th propagator is $D_j = p_j^2 - m_j^2$ with $p_j$ being the linear combination of loop and external momenta and $m_j$ the mass of corresponding propagator. A specific set of $D_j$ is called a propagator family. Normally, we can directly use FIESTA+ParInt program \cite{Alexander-8,Elise-2} to evaluate the MI in the physical region, but some of the principal integrals will be difficult to improve accuracy and the calculation is very time consuming. In the calculation of MIs, we firstly adopt the FIESTA+ParInt program using the sector decomposition method to get the values of the MIs in the non-physical region, where the convergence of the integral functions is faster and the MIs can be calculated efficiently with very high precision. Secondly, the obtained results serve as initial conditions of a suitable set of differential equations built upon all the MIs, and then the values of all MIs in physical region can be evaluated through the numerical integration of the differential equations \cite{Mandal,odeint}.

\par
Since the energy scale $f$ is constrained to be several TeV or even higher\cite{Reuter}, we omit the terms in couplings with order of $\mathcal{O}(v^2/f^2)$ (see Appendix A). Throughout our calculations we adopt the unitary gauge, and neglect the masses of electron, muon and light-quarks ($u,\, d,\, s$) due to their exceedingly tiny Yukawa couplings. Generally, the amplitude for $Z_H \rightarrow H^0\gamma$ at any order can be expressed as
\begin{eqnarray}
\label{Amplitude}
\mathcal{M}(Z_H \rightarrow H^0\gamma)
=\mathcal{M}^{\mu \nu}\epsilon_\mu(q)\epsilon_\nu(k_1),
\end{eqnarray}
where $q$ and $k_1$ are the four-momenta of $Z_H$ and $\gamma$, respectively. The matrix element $\mathcal{M}^{\mu \nu}$ can be written as
\begin{eqnarray}
\label{AmpStructure}
\mathcal{M}^{\mu\nu}
=
\mathcal{A}g^{\mu\nu}
+ \mathcal{B}\hat{q}^{\nu}\hat{k}_1^{\mu}
+ \mathcal{C}q_{\alpha}k_{1\beta}\epsilon^{\mu\nu\alpha\beta}
+ \mathcal{D}q^{\mu}k_1^{\nu}
+ \mathcal{E}q^{\mu}q^{\nu}
+ \mathcal{F}k_1^{\mu}k_1^{\nu},
\end{eqnarray}
where $\hat{k}_1= \dfrac{k_1}{m_{Z_H}}$ and $\hat{q} = \dfrac{q}{m_{Z_H}}$. As we know, the matrix element should satisfy the Ward identity, i.e., $k_{1\nu}\mathcal{M}^{\mu\nu}=0$, thus $\mathcal{E} = 0$ and $\mathcal{B}=\dfrac{2 m_{Z_H}^2}{(m_H^2-m_{Z_H}^2)} \mathcal{A}$. Furthermore, the coefficients $\mathcal{D}$ and $\mathcal{F}$ have no contribution to $|\mathcal{M}|^2$. Then we only consider the first three terms of the right side of $\mathcal{M}^{\mu \nu}$ in our calculation, i.e.,
\begin{eqnarray}
\label{ampW}
\mathcal{M}^{\mu \nu}
=
\mathcal{A} g^{\mu \nu}+ \mathcal{B} \hat q^\nu \hat k_1^\mu +\,\mathcal{C}q_{\alpha}k_{1\,\beta}\varepsilon^{\mu\nu\alpha\beta},
\end{eqnarray}
and the decay width for $Z_H \rightarrow H^0\gamma$ is obtained as
\begin{eqnarray}
\label{decaywidth}
\Gamma(Z_H \rightarrow H^0 \gamma)
=
\dfrac{1}{3}\frac{m_{Z_H}^2-m_{H}^2}{m_{Z_H}^3}\left[  \frac{\mathcal{A}^2}{8\pi}+\frac{(m_{Z_H}^2-m_{H}^2)^2}{32\pi}\mathcal{C}^2 \right].
\end{eqnarray}

\subsection{Leading-order amplitude}
\par
The leading-order (LO) contributions to the decay width of the $Z_H \rightarrow H^0\gamma$ process in the LHM have been comprehensively described in Ref.\cite{Aranda}. In this work we are going to evaluate the NLO QCD corrections to this decay process, and thus should calculate the LO amplitude at first. The LO one-loop Feynman diagrams can be divided into two sets of graphs: (1) triangle loop diagrams, and (2) tadpole and self-energy loop diagrams. Since the $T-\bar{T}-Z_H$ gauge coupling is at the $\mathcal{O}(v^2/f^2)$, it's reasonable to omit the pure $T$-quark triangle diagram.
\begin{figure}[htbp]
\begin{center}
\includegraphics[scale=0.45]{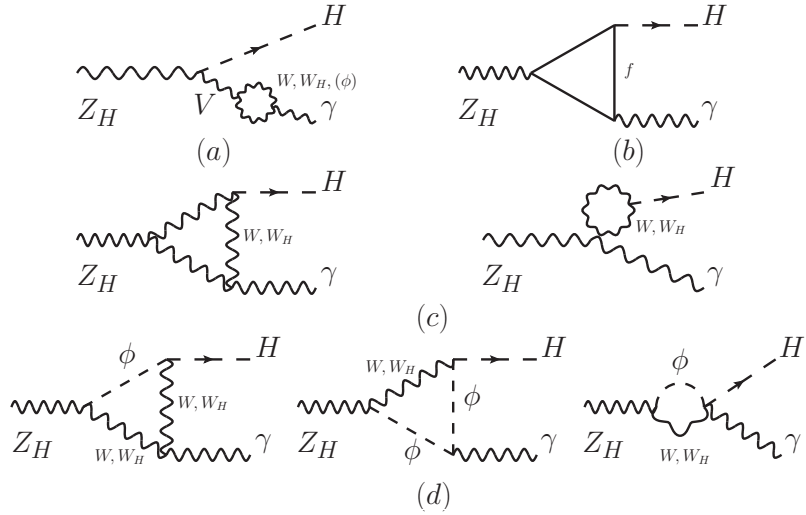}
\caption{Representative one-loop Feynman diagrams for $Z_H \rightarrow H^0\gamma$, where $f= \tau,\, c,\, b,\, t,\, t\text{-}T$, $V=Z,\, Z_H,\, A_H$, and $\phi$ denotes charged scalars.}
\label{fig1}
\end{center}
\end{figure}

\par
We depict some representative triangle one-loop Feynman diagrams which contribute to the LO decay width of $Z_H \rightarrow H^0\gamma$ in Figs.\ref{fig1}(a-d). Fig.\ref{fig1}(a) shows some self-energy diagrams of one external line. Fig.\ref{fig1}(b) represents the triangle loop diagrams which are mediated by massive charged fermions $f=\tau,\, c,\, b,\, t,\, t\text{-}T$). ($f=t\text{-}T$ represents the triangle loop diagrams with $t\text{-}T$ mixing.) In Fig.\ref{fig1}(c) the triangle graphs are actually mediated by the SM and new heavy charged gauge bosons and the mixing of these two types of particles. In Fig.\ref{fig1}(d) the typical loop graphs are induced by scalar and scalar plus gauge boson loops. Our calculation shows that the contribution from tadpole and self-energy diagrams vanishes. Then from all the relevant one-loop Feynman diagrams and the Feynman rules (Some of the relevant LHM couplings are listed in Appendix A) and using Eq.(\ref{ampW}) we can get the one-loop matrix element $\mathcal{M}_{LO}^{\mu \nu}$ as
\begin{eqnarray}
\label{ampW-1}
\mathcal{M}_{LO}^{\mu \nu}
=
\mathcal{A}_{LO} g^{\mu \nu}+ \mathcal{B}_{LO}  \hat q^\nu\hat k_1^\mu+\,\mathcal{C}_{LO}k_{1\,\alpha}q_{\beta}\varepsilon^{\mu\nu\alpha\beta}.
\end{eqnarray}

\par
In order to make comparison for the analytical expressions of the form factor coefficients with those in Ref.\cite{Aranda}, we follow the LO analysis in Ref.\cite{Aranda} and present the explicit amplitude expressions in Appendix B. All the form factor coefficients $\mathcal{A}_{LO}$, $\mathcal{B}_{LO}$ and $\mathcal{C}_{LO}$ are expressed in terms of Passarino-Veltman scalar functions, which are defined same as in Ref.\cite{Denner}. Furthermore, we divide each of the form factor coefficients, $\mathcal{A}_{LO}$ and $\mathcal{B}_{LO}$, into three parts contributed by different diagram sets as \footnote{The nonzero contribution to the form factor $\mathcal{C}_{LO}$ is only from the $t$-$T$ mixing quark triangle diagrams.}
\begin{eqnarray}\label{formfactor}
\mathcal{A}_{LO} &=&
\sum\limits_{f=\tau,c,b,t}^{t-T}
\mathcal{A}^{LO}_{f}
+
\sum\limits_{i=1}^{3}\mathcal{A}^{LO}_{G_{i}}
+
\sum\limits_{i=1}^{2}\mathcal{A}^{LO}_{S_{i}},
\nonumber \\
\mathcal{B}_{LO} &=&
\sum\limits_{f=\tau,c,b,t}^{t-T}
\mathcal{B}^{LO}_f
+
\sum\limits_{i=1}^{3}\mathcal{B}^{LO}_{G_{i}}
+
\sum\limits_{i=1}^{2}\mathcal{B}^{LO}_{S_{i}},
~~~~~~
\mathcal{C}_{LO} = \mathcal{C}^{LO}_{t\text{-}T},
\end{eqnarray}
where $f$ runs over $\tau,\, c,\, b,\, t$ and $t\text{-}T$ mixing in the LHM, $G_i$ symbolizes charged gauge bosons ($W$, $W_H$, and $W\text{-}W_H$ mixing), and $S_i$ denotes charged scalars. After our calculation we find that our expressions for the LO amplitude coefficients have some differences compared with the corresponding ones in Ref.\cite{Aranda}. Accordingly, we provide the explicit expressions for the one-loop form factor coefficients appeared in Eq.(\ref{formfactor}) in Appendix B.

\subsection{NLO QCD corrections}
\par
\begin{figure}[htbp]
\begin{center}
\includegraphics[scale=0.45]{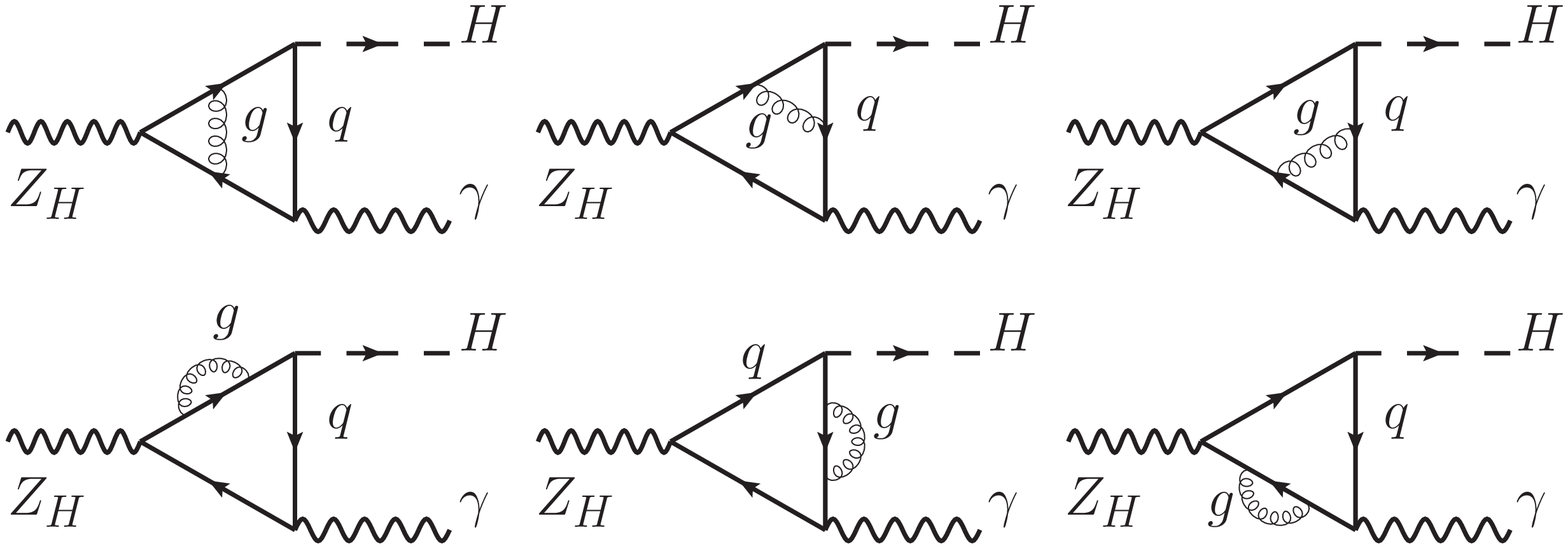}
\caption{Representative generic two-loop Feynman diagrams related to the NLO QCD corrections to the $Z_H \rightarrow H^0\gamma$ decay.}
\label{fig2}
\end{center}
\end{figure}

The $\mathcal{O}(\alpha_{{\rm ew}}^3\alpha_s)$ contribution to the decay width is from the interference between one-loop and QCD two-loop amplitudes for the decay channel $Z_H \rightarrow H^0\gamma$. The two-loop correction includes all the contributions from the generic two-loop Feynman diagrams shown in Fig.\ref{fig2} which are based on the heavy quark one-loop triangle diagrams in Fig.\ref{fig1}(a) and induced by attaching one gluon propagator to the heavy quark lines in every possible way. We express the unrenormalized two-loop amplitude, $\mathcal{M}_{2-loop}$, analytically by means of a number of independent MIs.
\begin{figure}[htbp]
\begin{center}
\includegraphics[scale=0.45]{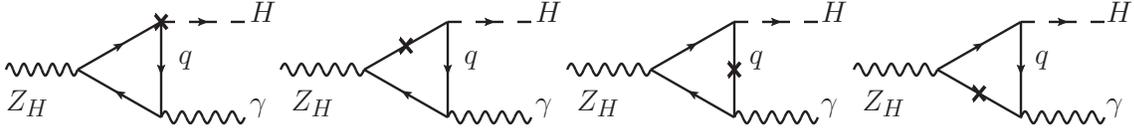}
\caption{Representative counterterm diagrams for $Z_H \rightarrow H^0\gamma$, where the crosses signify the NLO QCD counterterms for $q\bar{q}H$ $(q=c,\, b,\, t,\, t\text{-}T)$ vertices and quark propagators.}
\label{fig3}
\end{center}
\end{figure}

\par
The top family, corresponding to $q = t$ in Fig.\ref{fig2}, can be reduced to 31 MIs by adopting IBP technique. For example, a typical MI of the top family with 7 independent propagators is
\begin{eqnarray}
G(0,0,0,0,1,0,1)
\equiv
\int \frac{d^dl_1}{(2\pi)^d} \frac{d^dl_2}{(2\pi)^d} \frac{1}{(l_1^2-m_t^2)(l_2^2-m_t^2)}
=
\left[
-im_t^2\frac{\Gamma(\epsilon-1)}{16\pi^2} \left(\frac{4\pi\mu^2}{m_t^2}\right)^\epsilon
\right]^2,
\end{eqnarray}
where $l_{1,2}$ are the loop momenta and $\mu$ is the mass scale of dimensional regularization. The bottom and charm families, which correspond to $q = b$ and $q = c$ in Fig.\ref{fig2}, can be easily obtained from the top family by performing the replacements of $m_t \rightarrow m_b$ and $m_t \rightarrow m_c$, respectively. For $q=t\text{-}T$ in Fig.\ref{fig2}, there are two families and each family can be reduced to 35 MIs. It is obvious that the MIs of the two families can be obtained from each other by performing the exchange between $m_t$ and $m_T$ in all propagators.

\par
The NLO QCD bare amplitude $\mathcal{M}_{2-loop}$ has to be renormalized to remove the UV divergence. We choose the dimensional regularization in our calculation, and adopt the on-shell (OS) scheme \cite{Bernreuther-1} in handling the renormalization of quark-masses and Yukawa couplings. We note that there is no requirement for the renormalization for the relevant weak gauge couplings except the renormalization of the quark mass in Yukawa coupling, because the two-loop amplitude is the LO in $\alpha_s$. Actually, the QCD NLO amplitude renormalization for this decay channel is implemented by the charm-, bottom-, top- and $T$-quark mass renormalization for relevant Yukawa couplings, i.e., $c\bar{c}H$, $b\bar{b}H$, $t\bar{t}H$, $T\bar{T}H$ and $t\bar{T}H$ couplings. They are directly related to $\delta m_c$, $\delta m_b$, $\delta m_t$ and $\delta m_T$, and the counterterms for those couplings can be expressed as
\begin{eqnarray}
   \delta G_{f\bar{f}H}=- i \frac{\delta m_f}{v} \left[ 1 - \frac{1}{2} s_0^2 + \frac{v}{f} \frac{s_0}{\sqrt{2}}-\frac{2v^2}{3f^2}\right],~~~(f=\tau,\, c,\, b),  \label{GffH}
\end{eqnarray}
\begin{eqnarray}
   \delta G_{t\bar{t}H}=- i \frac{\delta m_t}{v} \left[ 1 - \frac{1}{2} s_0^2 + \frac{v}{f} \frac{s_0}{\sqrt{2}} - \frac{2 v^2}{3 f^2} +\frac{v^2}{f^2}c_{\lambda}^2 \left(1+c_{\lambda}^2\right) \right],       \label{GttH}
\end{eqnarray}
\begin{eqnarray}
   \delta G_{T\bar{T}H}=-i \frac{\delta m_T}{v}c_{\lambda}^2 \left(1+c_{\lambda}^2\right)\frac{v^2}{f^2}.  \label{GTTH}
\end{eqnarray}
For the counterterm of $\delta G_{t\bar{T}H}$, we have
\begin{eqnarray}
   \delta G_{t\bar{T}H}=\frac{\delta m_t}{v}\frac{v}{f}\left(1+c_{\lambda}^2\right)P_R+\frac{\delta m_T}{v}\frac{v}{f}c_{\lambda}^2 P_L.    \label{GtTH}
\end{eqnarray}
We write the NLO QCD renormalized amplitude $\Delta \mathcal{M}_{NLO}$ as
\begin{eqnarray}
\Delta \mathcal{M}_{NLO} =  \mathcal{M}_{2-loop} + \mathcal{M}_{CT},
\label{amp-NLO}
\end{eqnarray}
where $\mathcal{M}_{2-loop}$ and $\mathcal{M}_{CT}$ are the amplitudes contributed by two-loop diagrams and its corresponding NLO QCD counterterms separately. The counterterm amplitude $\mathcal{M}_{CT}$ comes from the contributions of counterterm diagrams shown in Fig.\ref{fig3}. We divide the total counterterm amplitude into four groups, i.e., $\mathcal{M}^q_{CT}$ ($q=c,\, b,\, t,\, t\text{-}T$), which are $c$-, $b$-, $t$-quark and $t\text{-}T$ mixing triangle loop diagram groups, respectively. Each group has four diagrams with a cross marked on one propagator or vertex as shown in Fig.\ref{fig3}. The total counterterm amplitude from Fig.\ref{fig3} can be expressed as
\begin{eqnarray}
\mathcal{M}_{CT}
=
\sum_{q=c,b,t}^{t-T}
\mathcal{M}^q_{CT}
=
\left[
\frac{\delta m_b}{m_b}\big( \mathcal{M}^b_{LO} + \mathcal{G}^b \big)
+
( b \rightarrow c )
\right]
+
\frac{\delta m_t}{m_t} \mathcal{H} +\frac{\delta m_T}{m_T} \mathcal{K},
~~
\label{amp-CT1}
\end{eqnarray}
where $\mathcal{M}^b_{LO}$ is the LO amplitude for the $b$-quark one-loop triangle diagrams, $\dfrac{\delta m_b}{m_b} \mathcal{M}^b_{LO}$ and $\dfrac{\delta m_b}{m_b} \mathcal{G}^b$ are the contributions induced by the NLO QCD counterterms for $b\bar{b}H$ vertex and $b$-quark propagator, i.e., the contributions from the first and the last three diagrams in Fig.\ref{fig3} for $q=b$, respectively. $\mathcal{H}$ and $\mathcal{K}$ can be obtained by computing the $t$-quark and $t\text{-}T$ mixing triangle diagrams in Fig.\ref{fig3}. In the OS scheme the heavy quark mass counterterm is given by \cite{Bernreuther-1}
\begin{eqnarray}
  \delta m_{q}
   =
   -m_q \frac{\alpha_{s}(\mu)}{\pi}C(\epsilon)\left( \frac{\mu^2}{m_{q}^2}\right)^{\epsilon} \frac{C_{F}}{4}\frac{(3-2\epsilon)}{\epsilon(1-2\epsilon)}
   =
   -\frac{3C_F}{4}\frac{\alpha_{s}(\mu)}{\pi}\left( \frac{1}{\epsilon}-\gamma_E+\frac{4}{3}+\ln\frac{4\pi \mu^2}{m_q^2}  \right)
   \\
   (q=c,\, b, \, t, \, T), \nonumber
\end{eqnarray}
where $\epsilon = \dfrac{4-d}{2}$, $\gamma_E$ is the Euler constant, $C_F=\dfrac{3}{4}$, $C(\epsilon)=(4\pi)^{\epsilon}\Gamma(1+\epsilon)$, and $\mu$ is the mass scale of dimensional regularization. Finally, the total renormalized amplitude is expressed in terms of a certain number of independent MIs, and their numerical calculations are performed by adopting the FIESTA+ParInt program combined with the differential equation method.

\section{Numerical results and discussion}
\par
In this section, we present some numerical results of the LO and NLO QCD corrected $Z_H \rightarrow H^0\gamma$ decay widths in the LHM without $T$-parity. In the numerical calculation, we ignore the masses of electron, muon and light-quark masses i.e., $m_e = m_{\mu} = m_u = m_d = m_s = 0$, and take the other relevant SM input parameters as follows \cite{Tanabashi}
\begin{eqnarray}
&&
m_W = 80.379~\text{GeV},~~~~ \quad m_Z = 91.1876~\text{GeV},~~~~ \quad m_H = 125.18~\text{GeV},
\nonumber \\
&&
m_{\tau} = 1.77686~\text{GeV},~~~~ \quad m_c = 1.67~\text{GeV},~~~~ \quad m_b = 4.78~\text{GeV},~~~~ \quad m_t = 173.1~\text{GeV},
\nonumber \\
&&
\alpha_{ew} = 1/137.035999139,~~~~G_F = 1.16638\times10^{-5}~\text{GeV}^{-2}.
\end{eqnarray}
The VEV in the SM, $v_{SM}$, can be got as $v_{SM}=(\sqrt{2}G_{F})^{-1/2} \approx 246~\text{GeV}$, and one of the VEVs in the LHM, $v$, which triggers the electroweak symmetry breaking gets a modification up to the $\mathcal{O}(v_{SM}^2/f^2)$ as \cite{WangL}
\begin{eqnarray}
v = v_{SM}\left[ 1-\frac{v_{SM}^2}{f^2}\left(1-\frac{5}{24}+\frac{x^2}{8}\right) \right].
\end{eqnarray}
The strong coupling constant $\alpha_s(\mu)$ is obtained by the expression in the $\overline{\text{MS}}$ scheme up to the two-loop order. We applied the Mathematica package RunDec \cite{Chetyrkin} to evolve the strong coupling constant $\alpha_s$ up to scale $\mu= m_{Z_H}$.

\par
In the LHM there are five independent input parameters in addition to the SM input parameters, which are chosen as $f$, $c$, $c'$, $x$, and $\lambda_1/\lambda_2$. In our numerical calculation, we take $f = 3,\, 4~\text{TeV}$, $x=0$, $c^{\prime}=1/\sqrt{2}$, $\lambda_1/\lambda_2=1$, and $c$ parameter varying from $0.1$ to $0.6$.

\par
\begin{figure}[htbp]
\begin{center}
\includegraphics[scale=0.3]{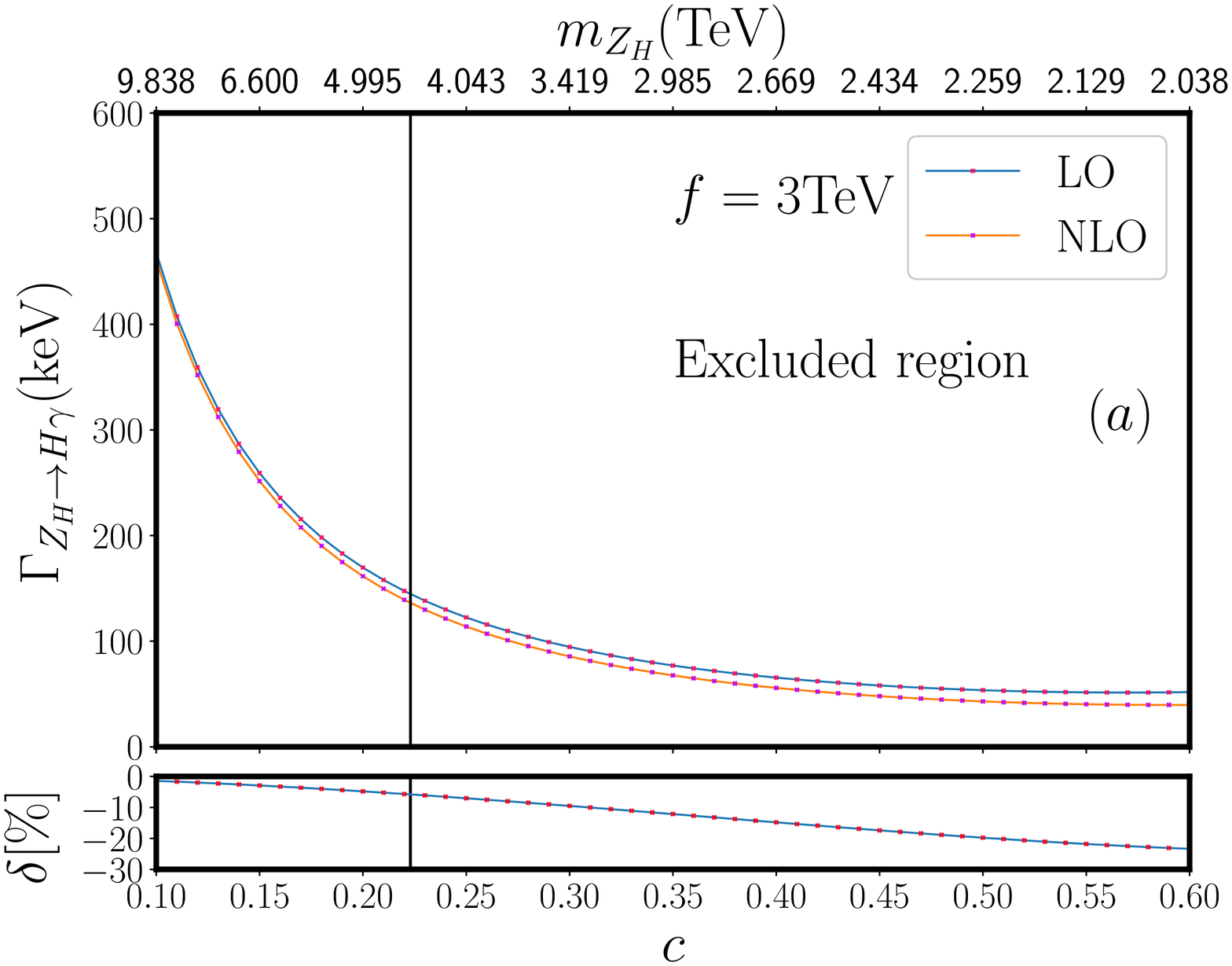}
\includegraphics[scale=0.3]{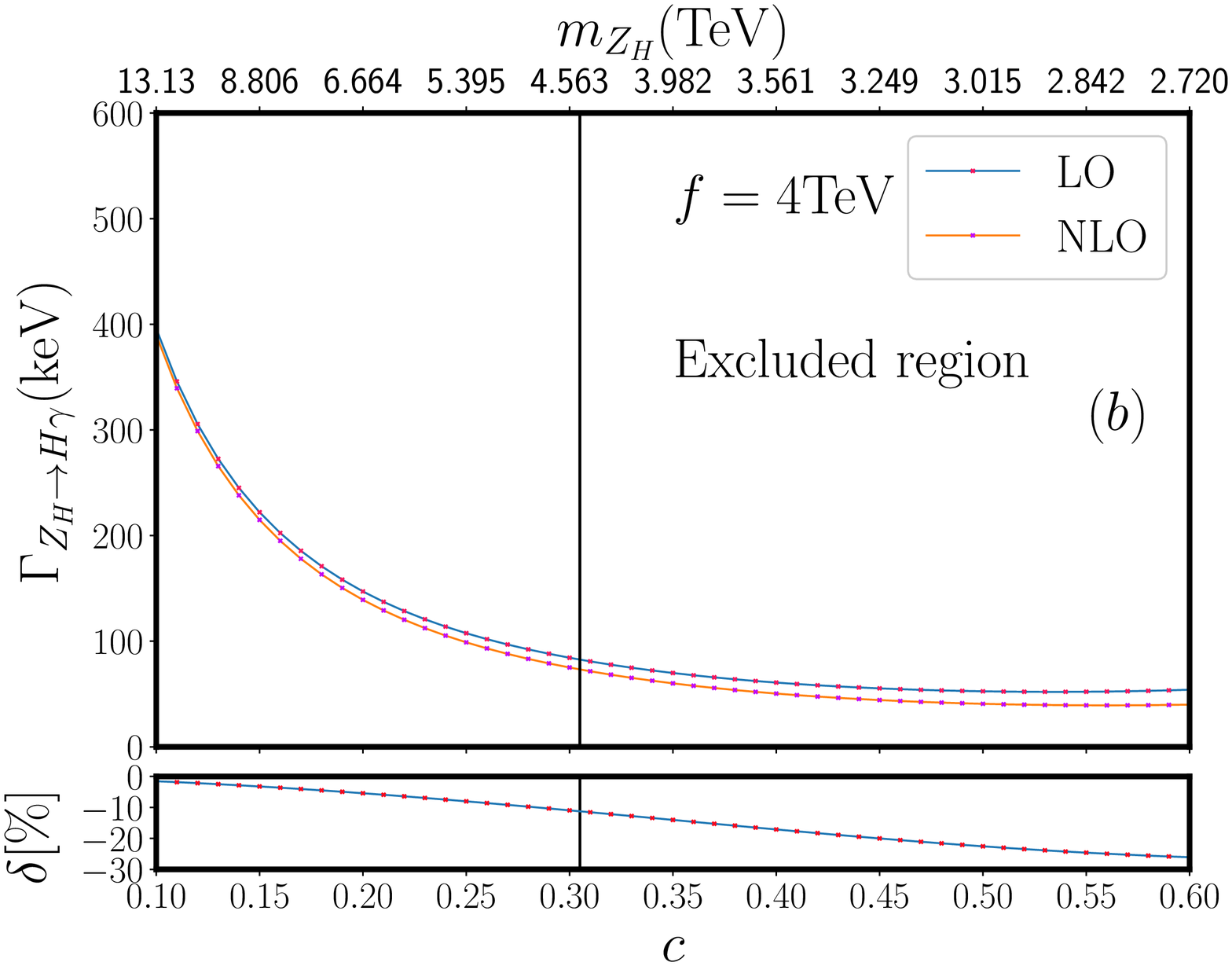}
\caption{LO and NLO QCD corrected decay widths and the corresponding relative corrections for the $Z_H \rightarrow H^0\gamma$ process versus $m_{Z_H}$ (or parameter $c$) for (a) $f = 3~\text{TeV}$ and (b) $f = 4~ \text{TeV}$.}
\label{fig4}
\end{center}
\end{figure}

In Figs.\ref{fig4}(a) and (b), we depict the LO and NLO QCD corrected decay widths of the $Z_H \rightarrow H^0\gamma$ decay as functions of the $Z_H$ mass (or parameter $c$), for $f = 3,\, 4~\text{TeV}$ separately. Recently, ATLAS experiment provides a lower limit of up to $4.5~\text{TeV}$ on the mass of heavy neutral vector boson $Z_H$ \cite{Atlas-1}, hence in these plots we mark out the present excluded regions which are beyond the most recent experimental constraints on the parameters space. We see from the two figures that the plotted experiment permitted region for $Z_H$ mass is $m_{Z_H} \in [4.5~\text{TeV},~ 9.838~\text{TeV}]$ (corresponding to $c \in [0.1,~ 0.223]$) for $f = 3~\text{TeV}$, and $m_{Z_H} \in [4.5~\text{TeV},~ 13.13~\text{TeV}]$ (i.e., $c \in [0.1,~ 0.305]$) for $f=4~\text{TeV}$ correspondingly. Fig.\ref{fig4}(a) for $f=3~\text{TeV}$ shows that when the parameter $c$ increases from $0.1$ to $0.22$ in the experiment allowed range, the LO (NLO) decay width decreases from $467.585~\text{keV}$ ($460.878~\text{keV}$) to $147.573~\text{keV}$ ($139.190~\text{keV}$).
While Fig.\ref{fig4}(b) for $f=4~\text{TeV}$ tells us that the LO (NLO) decay width decreases from $395.925~\text{keV}$ ($389.752~\text{keV}$) to $84.342~\text{keV}$ ($75.099~\text{keV}$) with the increment of parameter $c$ from $0.1$ to $0.3$. These data read off from Figs.\ref{fig4}(a) and (b) correspond to different values of the LHM parameter set ($c$ and $f$), which are in the most recent experiment permitted regions, are also listed in Table \ref{table-1}. The corresponding NLO QCD relative corrections to the $Z_H \rightarrow H^0\gamma$ decay width are presented in the table too. We can see from Figs.\ref{fig4}(a, b) and Table \ref{table-1} that the NLO QCD correction always diminishes the decay width of the $Z_H \rightarrow H^0\gamma$ process in our chosen parameter space, and the NLO QCD relative correction can reach $-11.0\%$ for $f=4~\text{TeV}$ and $c=0.30$. It shows that the QCD two-loop correction is very significant and should be included in the precision prediction of the decay width.
\begin{table}[htbp]
\begin{center}
\renewcommand\arraystretch{1.5}
\begin{tabular}{c|c|c|c|c}
\hline
\hline
$f~ \text{(TeV)}$  & \multicolumn{2}{c|}{3} &\multicolumn{2}{c}{4} \\
\hline
$c$  & $0.1$ & $0.22$ & $0.1$ & $0.3$ \\
\hline
\hline
$\Gamma^{\text{LO}}~  \text{(keV)}$  & $467.585$ & $147.573$  & $395.925$ & $84.342$ \\
\hline
$\Gamma^{\text{NLO}}~ \text{(keV)}$  & $460.878$ & $139.190$  & $389.752$ & $75.099$ \\
\hline
$\delta~ (\%)$                       & $-1.43$   & $-5.68$    & $-1.56$   & $-11.0$  \\
\hline
\hline
\end{tabular}
\caption{LO and NLO QCD corrected decay widths for $Z_H \rightarrow H^0\gamma$ and the corresponding relative QCD corrections for some typical values of $c$ and $f$.}
\label{table-1}
\end{center}
\end{table}

\par
\begin{figure}[htbp]
\begin{center}
\includegraphics[scale=0.3]{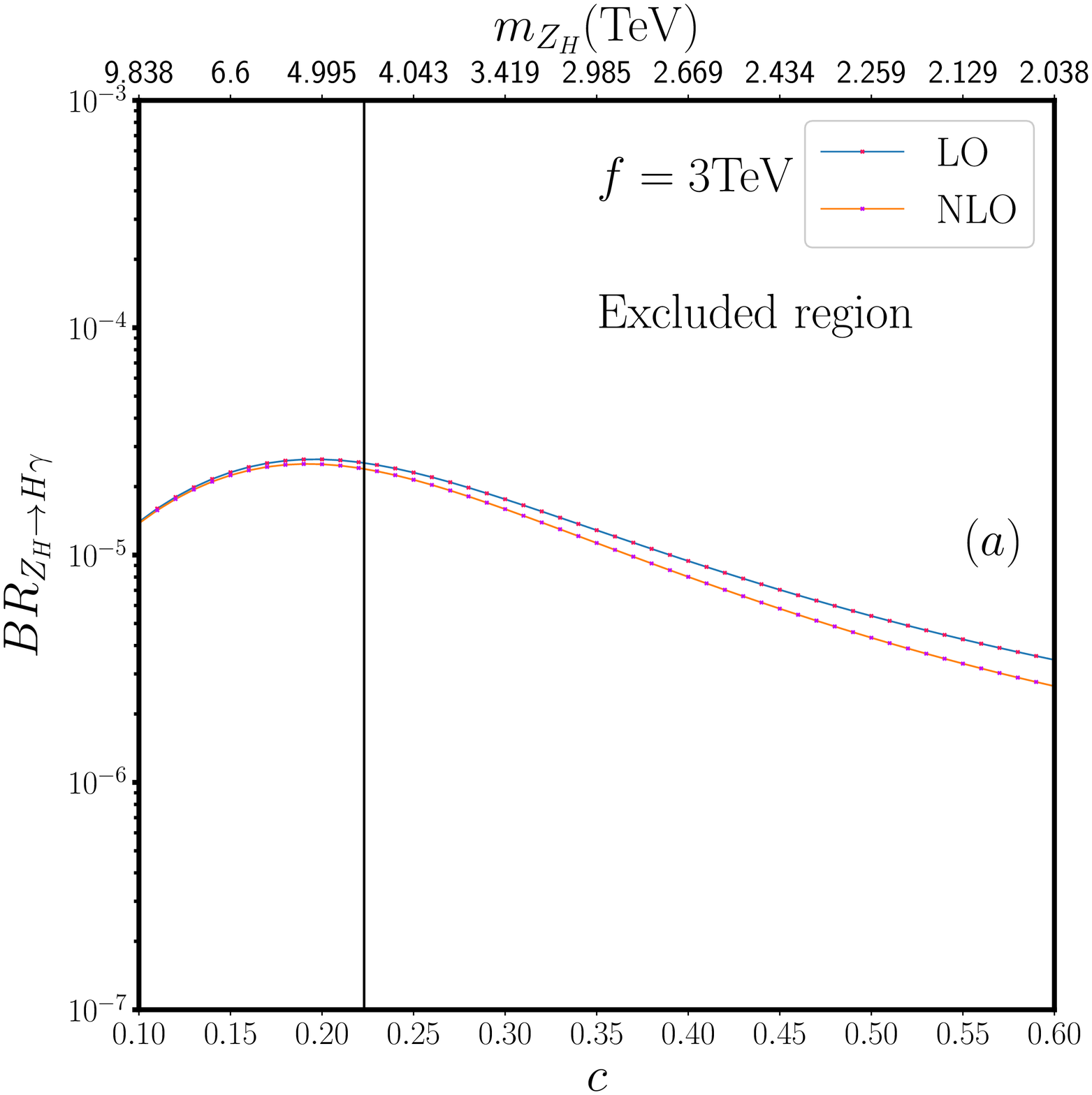}
\includegraphics[scale=0.3]{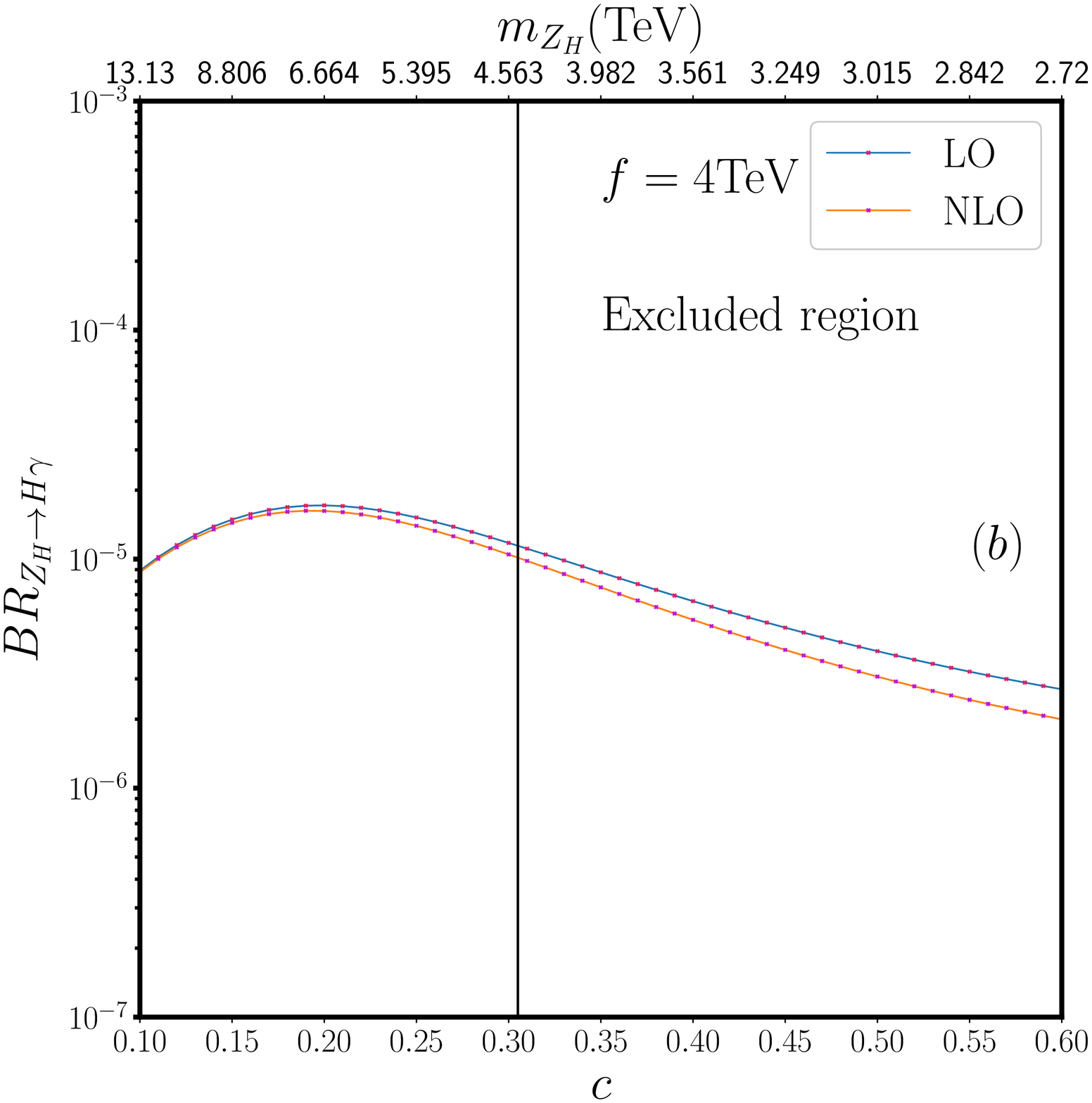}
\caption{LO and NLO QCD corrected branching ratio for $Z_H \rightarrow H^0\gamma$ versus $m_{Z_H}$ (or parameter $c$) for (a) $f = 3~\text{TeV}$ and (b) $f = 4~ \text{TeV}$.}
\label{fig5}
\end{center}
\end{figure}

In Ref.\cite{Aranda}, the authors calculated the branching ratio of the decay channel $Z_H \rightarrow H^0\gamma$ only at the LO. For comparison, we also depict the LO as well as the NLO QCD corrected branching ratio of $Z_H \rightarrow H^0\gamma$ as a function of $c$ in Figs.\ref{fig5}(a) and (b) for $f = 3$ and $4~\text{TeV}$, respectively, where $c$ varies in the range of $0.1 < c < 0.6$. The total decay width of $Z_H$ is calculated by using the analytical expressions for the partial decay widths of the dominant decay channels of $Z_H$\cite{ZHwidth}. We can see that the curves for LO branching ratio in Figs.\ref{fig5}(a) and (b) behave similarly as the corresponding ones in Ref.\cite{Aranda}, but have different branching ratio values. As we know, if we only consider the contribution from the $W$-boson and SM fermion loops, the LO decay width of $Z_H \rightarrow H^0\gamma$ can be obtained from the analytical expression for the LO decay width of $Z \rightarrow H^0\gamma$ in Ref.\cite{SMZHr} by rescaling some coupling strengths and performing the replacement of $m_Z \rightarrow m_{Z_H}$. To check the correctness of our LO calculation, we compute the contribution from the $W$-boson and SM fermion loops to the decay width of $Z_H \rightarrow H^0\gamma$, and find that our numerical result is coincident with that obtained from the decay width of $Z \rightarrow H^0\gamma$ by performing relevant replacements within the calculation error.

\par
Now let's discuss the contributions from various groups of diagrams. Firstly, we separate the total contribution to the LO decay width of the $Z_H \rightarrow H^0\gamma$ decay ($\Gamma^{\text{LO}}$) into three origins: (1) the amplitude of all the boson one-loop diagrams, $\mathcal{M}_{LO}^B=\sum\limits_{i}\mathcal{M}_{LO}^i$, where the superscript $i$ runs over $\Phi$, $W^{\pm}$, $W_H^{\pm}$ and $W^{\pm}$-$W_H^{\pm}$ mixing one-loop triangle diagrams, (2) the amplitude of all the fermion one-loop diagrams, $\mathcal{M}_{LO}^F=\sum\limits_{f}\mathcal{M}_{LO}^f$ with $f$ running over $\tau,\, c,\, b,\, t$ and $t\text{-}T$-mixing triangle diagrams, and (3) the interference between the above two amplitudes. Then we can write the decay width as
\begin{equation}
\Gamma^{\text{LO}}=\Gamma^{\text{LO}}_{BB}+\Gamma^{\text{LO}}_{BF}+\Gamma^{\text{LO}}_{FF}.
\end{equation}
The values of the partial decay widths from above three components are listed in Table \ref{table-2}. From this table we see clearly that $\Gamma^{\text{LO}}_{BB}$ provides most of the contributions and has to be taken into account. The interference between the amplitudes of $\mathcal{M}_{LO}^{B}$ and $\mathcal{M}_{LO}^{F}$, $\Gamma^{\text{LO}}_{BF}$, gives the contribution of one order of magnitude smaller than $\Gamma^{\text{LO}}_{BB}$, while the contribution component $\Gamma^{\text{LO}}_{FF}$ is about one order smaller than $\Gamma^{\text{LO}}_{BF}$.
\begin{table}[htbp]
\begin{center}
\renewcommand\arraystretch{1.5}
\begin{tabular}{c|c|c}
\hline
\hline
 Partial decay width  & $f=3~ \text{TeV},~ c=0.2$ & $f=4~ \text{TeV},~ c=0.3$ \\
\hline
$\Gamma^{\text{LO}}_{BB}~ \text{(keV)}$  & $146.258$ & $56.770$ \\
\hline
$\Gamma^{\text{LO}}_{BF}~ \text{(keV)}$  & $20.103$  & $19.823$ \\
\hline
$\Gamma^{\text{LO}}_{FF}~ \text{(keV)}$  & $3.360$   & $7.749$  \\
\hline
\hline
\end{tabular}
\caption{LO contributions to the decay width of $Z_H \rightarrow H^0 \gamma$. $\Gamma^{\text{LO}}_{BB}$ and $\Gamma^{\text{LO}}_{FF}$ are the contributions induced by the boson and fermion loops, respectively, while $\Gamma^{\text{LO}}_{BF}$ represents the interference between the boson and fermion loop amplitudes.}
\label{table-2}
\end{center}
\end{table}

\par
We list some typical values of the two-loop QCD corrections to the decay width in Table \ref{table-3}. The correction component $\Delta \Gamma^{\text{NLO}}_{Bq}$ ($\Delta \Gamma^{\text{NLO}}_{Fq}$) describes the contribution from the interference between the boson (fermion) one-loop amplitude $\mathcal{M}_{\text{LO}}^B$ ($\mathcal{M}_{\text{LO}}^F$) and the amplitude $\mathcal{M}_{\text{NLO}}^q$ for the diagrams with $q$-quark in two-loop. The superscript $q$ represents the possible quark ($c,\, b,\, t$ or $t\text{-}T$) in QCD two-loop (shown in Fig.\ref{fig2}). We can see that the most dominant NLO QCD correction to the decay width is $\Delta \Gamma^{\text{NLO}}_{Bt}$, and $\Delta \Gamma^{\text{NLO}}_{Ft}$ is the second largest NLO QCD contribution. The NLO QCD contributions$\Delta \Gamma^{\text{NLO}}_{Bb}$, $\Delta \Gamma^{\text{NLO}}_{Fb}$, $\Delta \Gamma^{\text{NLO}}_{Bc}$ and $\Delta \Gamma^{\text{NLO}}_{Fc}$ are about three orders of magnitude smaller than $\Delta \Gamma^{\text{NLO}}_{Ft}$, and thus can be neglected in the case of our chosen LHM parameter space region. As shown in Eq.(\ref{eq-1}) in Appendix B, the coefficients in $\mathcal{M}_{LO}^{t-T}$ for the one-loop $t\text{-}T$ mixing triangle diagrams have the values as $\mathcal{A}^{LO}_{t-T} = \mathcal{B}^{LO}_{t-T} = 0$, and $\mathcal{C}^{LO}_{t-T}$ is nonzero. Therefore, $\Delta \Gamma^{\text{NLO}}_{F,t-T}$ is actually only contributed by the nonzero interference between the one-loop amplitude $\mathcal{M}_{LO}^{t-T}$ and the two-loop amplitude $\mathcal{M}_{NLO}^{t-T}$. From Table \ref{table-3} we can see that only $\Delta \Gamma^{\text{NLO}}_{F,t-T}$ has positive value, which is the third largest correction part among all the seven correction parts listed in the table. We can conclude that the top-induced two-loop contribution is the main source of the NLO QCD correction.
\begin{table}[htbp]
\begin{center}
\renewcommand\arraystretch{1.5}
\begin{tabular}{c|c|c}
\hline
\hline
 Partial decay width & $f=3~ \text{TeV},~ c=0.2$ & $f=4~ \text{TeV},~ c=0.3$    \\
\hline
$\Delta \Gamma^{\text{NLO}}_{Bt}~ \text{(keV)}$     & $-7.104$    & $-6.743$    \\
\hline
$\Delta \Gamma^{\text{NLO}}_{Bb}~ \text{(keV)}$     & $-0.00600$  & $-0.00587$  \\
\hline
$\Delta \Gamma^{\text{NLO}}_{Bc}~ \text{(keV)}$     & $-0.000857$ & $-0.000833$ \\
\hline
$\Delta \Gamma^{\text{NLO}}_{Ft}~ \text{(keV)}$     & $-1.086$    & $-2.481$    \\
\hline
$\Delta \Gamma^{\text{NLO}}_{Fb}~ \text{(keV)}$     & $-0.00450$  & $-0.0106$   \\
\hline
$\Delta \Gamma^{\text{NLO}}_{Fc}~ \text{(keV)}$     & $-0.00156$  & $-0.00367$  \\
\hline
$\Delta \Gamma^{\text{NLO}}_{F,t-T}~ \text{(keV)}$  & $0.0107$    & $0.00224$   \\
\hline
\hline
\end{tabular}
\caption{NLO QCD contributions to the decay width of $Z_H \rightarrow H^0\gamma$, $\Delta \Gamma^{\text{NLO}}_{iq}$, where $i=B,\, F$ and $q = c,\, b,\, t,\, t\text{-}T$, for some typical values of the LHM parameters $f$ and $c$.}
\label{table-3}
\end{center}
\end{table}

\par
\section{Summary}
In this work we investigate the $Z_H \rightarrow H^0 \gamma$ decay channel in the LHM without $T$-parity up to the $\mathcal{O}(\alpha_{{\rm ew}}^{3}\alpha_s)$. At the LO level we involve the contributions from the one-loop diagrams mediated by heavy fermions, scalars, gauge bosons, and the admixture of these later two type particles. We revisit analytically and numerically the LO decay width for $Z_H \rightarrow H^0\gamma$ and compared them with the previous work. In our calculation, we accomplish the two-loop evaluation by using the integration-by-parts identities for the reduction to master integrals. The numerical integration for the MIs is carried out by our developed program combining the FIESTA+ParInt package with the differential equations method. The LO and NLO QCD corrected decay widths are calculated by taking the LHM input parameters $f = 3,~4~\text{TeV}$ and $0.1<c<0.6$. We focus on the discussion of the numerical results of the decay width and NLO QCD correction by taking the LHM parameters within the recent experimental constraint region. We find that in the LHM parameter space region we considered, the NLO QCD correction is always negative and the top related QCD correction is the dominant contribution at the QCD NLO. For $f=4~\text{TeV}$ and $c=0.3$, the NLO QCD corrected decay width has the value of $75.099~\text{keV}$ and the NLO QCD relative correction can reach $-11.0\%$.

\vskip 5mm
\par
\noindent{\large\bf Acknowledgments} \\
This work is supported in part by the National Natural Science Foundation of China (Grants No. 11775211, No. 11535002), and the CAS Center for Excellence in Particle Physics (CCEPP).

\vskip 5mm
\section{Appendix }
\subsection{Appendix A: Relevant couplings}
\par
The Feynman rules of the couplings relevant to our work, can be read out from the Lagrangian shown in Eq.(\ref{L1}), which have been already provided in Ref.\cite{LH8,Buras}.  In the following we list some of the related LHM couplings in unitary gauge.
\begin{eqnarray}
\label{httLH-3}
G_{f\bar{f}H}^{LH} = - i \frac{m_f}{v} \left[ 1 - \frac{1}{2}s_0^2+\frac{v}{f} \frac{s_0}{\sqrt{2}}\right],~~~
(f=\tau,\, c,\, b),
\end{eqnarray}
\begin{eqnarray}
\label{httLH-2}
G_{t\bar{t}H}^{LH} = - i \frac{m_t}{v} \left[ 1 - \frac{1}{2}s_0^2+\frac{v}{f} \frac{s_0}{\sqrt{2}} - \frac{2v^2}{3f^2}+\frac{v^2}{f^2}c_\lambda^2\left(1+c_\lambda^2\right) \right],
\end{eqnarray}
\begin{eqnarray}
\label{httLH-4}
G_{T\bar{T}H}^{LH} = - i c_{\lambda}^2 \left( 1 + c_{\lambda}^2 \right) \frac{v}{f},
\end{eqnarray}
\begin{eqnarray}
\label{httLH-5}
G_{t\bar{T}H}^{LH} = \frac{m_t}{v}\frac{v}{f}\left(1+ c_{\lambda}^2\right)P_R+
\frac{m_T}{v}\frac{v}{f} c_{\lambda}^2 P_L,
\end{eqnarray}
\begin{eqnarray}
\label{httLH-6}
G_{\tau\bar{\tau}Z_H}^{V,LH} = -\frac{gc}{4s},~~~~~~~~~~G_{\tau\bar{\tau}Z_H}^{A,LH} = \frac{gc}{4s},
\end{eqnarray}
\begin{eqnarray}
\label{httLH-7}
G_{b\bar{b}Z_H}^{V,LH} = -\frac{gc}{4s},~~~~~~~~~~G_{b\bar{b}Z_H}^{A,LH} = \frac{gc}{4s},
\end{eqnarray}
\begin{eqnarray}
\label{httLH-8}
G_{c\bar{c}Z_H}^{V,LH} = \frac{gc}{4s},~~~~~~~~~~G_{c\bar{c}Z_H}^{A,LH} = -\frac{gc}{4s},
\end{eqnarray}
\begin{eqnarray}
\label{httLH-9}
G_{t\bar{t}Z_H}^{V,LH} = \frac{gc}{4s},~~~~~~~~~~G_{t\bar{t}Z_H}^{A,LH} = -\frac{gc}{4s},
\end{eqnarray}
\begin{eqnarray}
\label{httLH-10}
G_{T\bar{T}Z_H}^{V,LH} \sim \mathcal{O}(\frac{v^2}{f^2}),~~~~~~~~~G_{T\bar{T}Z_H}^{A,LH} \sim \mathcal{O}(\frac{v^2}{f^2}),
\end{eqnarray}
\begin{eqnarray}
\label{httLH-11}
G_{t\bar{T}Z_H}^{V,LH} = g c_{\lambda}^2 \frac{vc}{4fs},~~~~~~~~~~G_{t\bar{T}Z_H}^{A,LH} = -g c_{\lambda}^2 \frac{vc}{4fs},
\end{eqnarray}
where $P_{L,R}=(1\mp \gamma_5)/2$, $G^V$ and $G^A$ are the vector and axial-vector coupling constants shown as $i(G^V+G^A\gamma_5)\gamma^{\mu}$, and $s_0$ gives the mixing of Higgs fields, $s_0 \simeq 2 \sqrt{2}\dfrac{v^{\prime}}{v} =\dfrac{xv}{\sqrt{2}f} \sim \mathcal{O}(v/f)$.

\subsection{Appendix B: Amplitude coefficients}
\par
Here we provide the explicit formulas for the relevant form factor coefficients introduced in (\ref{formfactor}). For fermion loop the coefficients are given by
\begin{eqnarray}
\mathcal{A}^{LO}_f &=& \frac{N_f^c Q_f T_f^3}{16\pi^2} \frac{g^2 s_W c }{s(y_H-1)} m_f g_{f\bar{f}H}
\Big[2(B_a-B_b)+(y_H-1)(C_a(4y_f-y_H+1)+2) \Big],   \nonumber  \\
\mathcal{B}^{LO}_f &=& \frac{2}{(y_H-1)}\,\mathcal{A}_{f},   \nonumber \\
\mathcal{A}^{LO}_{t-T} &=& \mathcal{B}^{LO}_{t-T} = 0, \nonumber \\
\mathcal{C}^{LO}_{t-T} &=& \frac{1}{4\pi^2}\frac{g^2 s_W c}{s v}\frac{v^2}{f^2}c_\lambda^2 \Big[ y_t(c_\lambda^2+1)C_c-y_T c_\lambda^2 C_d \Big],
\label{eq-1}
\end{eqnarray}
where $f=\tau,\, c,\, b,\, t$, $N_c^{\tau} = 1$, $N^{c}_c=N^{b}_c=N^{t}_c=3$, and we define $y_{f}=m_{f}^2/m_{Z_H}^2$, $y_{H} = m_H^2/m_{Z_H}^2$ and $y_{T}=m_{T}^2/m_{Z_H}^2$. $Q_f$ is the charge of fermion, i.e., $Q_{\tau}=-1$, $Q_{c}=Q_{t}=2/3$ and $Q_{b}=-1/3$. $T_f^3$ denotes the third component of isospin: $T_{c}^3 = T_{t}^3 = 1$ and $T_b^3=T_\tau^3=-1$. The coupling constant $g_{\bar{f}fH}$ in the first expression of Eq.(\ref{eq-1}) is defined as $g_{f\bar{f}H} = i G_{f\bar{f}H}$ where the explicit expressions for $G_{f\bar{f}H}$ are given in Eqs.(\ref{httLH-3}-\ref{httLH-4}). $B_a,~ B_b,~ C_a,~ C_c$ and $C_d$ are defined as
\begin{eqnarray}
\label{eq-3}
B_a &=& B_0(m_H^2,m_f^2,m_f^2), \nonumber \\
B_b &=& B_0(m_{Z_H}^2,m_f^2,m_f^2), \nonumber \\
C_a &=& m_{Z_H}^2 C_0(m_H^2,m_{Z_H}^2,0,m_f^2,m_f^2,m_f^2), \nonumber \\
C_c &=& m_{Z_H}^2C_0(m_H^2,m_{Z_H}^2,0,m_t^2,m_T^2,m_t^2), \nonumber \\
C_d &=& m_{Z_H}^2C_0(m_H^2,m_{Z_H}^2,0,m_T^2,m_t^2,m_T^2),
\end{eqnarray}
where the integral functions $B_0$ and $C_0$ are the known Passarino-Veltman scalar functions.

\par
For the one-loop diagrams containing $W^\pm$ and $W_H^\pm$ bosons, the coefficients $\mathcal{A}_{G_{i}}$, $\mathcal{B}_{G_{i}}$ and $\mathcal{C}_{G_{i}}$ ($i=1,2,3$) are given by
\begin{eqnarray}
\mathcal{A}^{LO}_{G_{1}} &=& C_{G_{1}} \frac{1}{64\pi^2 (y_H-1)y_W^2}\Big[ (B_{G_{1a}}-B_{G_{1b}}) (y_H (1-2 y_W)+2 (1-6 y_W) y_W) \nonumber \\
&& -\, 2 C_{G_{1a}}\, y_W \left(y_H^2(1-6 y_W)+3 y_H \left(4 y_W^2+4 y_W-1\right)-12 y_W^2-6 y_W+2\right) \nonumber \\
&& +\, y_H^2 (1-2 y_W)+y_H \left(-12 y_W^2+4 y_W-1\right)+2 y_W (6 y_W-1)\Big],  \nonumber \\
\mathcal{B}^{LO}_{G_{1}} &=& \frac{2}{(y_H-1)}\,\mathcal{A}_{G_{1}},  \nonumber \\
\mathcal{C}^{LO}_{G_{1}} &=& 0,
\end{eqnarray}
where $C_{G_1} = -\dfrac{1}{2 f^2}\left[c g^4 s (c^2-s^2) s_W v^3\right]$, $y_{H} = m_H^2/m_{Z_H}^2$, $y_W = m_W^2/m_{Z_H}^2$,  and we define
\begin{eqnarray}
B_{G_{1a}} &=& B_0(m_H^2,m_W^2,m_W^2), \nonumber \\
B_{G_{1b}} &=& B_0(m_{Z_H}^2,m_W^2,m_W^2), \nonumber \\
C_{G_{1a}} &=& m_{Z_H}^2 C_0(m_H^2,m_{Z_H}^2,0,m_W^2,m_W^2,m_W^2).
\end{eqnarray}
Moreover, $\mathcal{A}^{LO}_{G_{2}}$ and $\mathcal{B}^{LO}_{G_{2}}$ can be obtained from $\mathcal{A}^{LO}_{G_{1}}$ and $\mathcal{B}^{LO}_{G_{1}}$ by performing the replacement of $m_W\to m_{W_H}$ and $C_{G_1}\to C_{G_2}$, where $C_{G_2}=-\dfrac{1}{ c s}\left[g^4 s_W v (c^2-s^2)\right]$.

\par
The coefficients $\mathcal{A}^{LO}_{G_{3}}$, $\mathcal{B}^{LO}_{G_{3}}$ and $\mathcal{C}^{LO}_{G_{3}}$ are concerned with the loop diagrams with the mixing between $W$ and $W_H$, and they are given by
\begin{eqnarray}
\mathcal{A}^{LO}_{G_3} &=& C_{G_3} \frac{1}{32\pi^2 (y_H-1) y_W y_{W_H}} \nonumber \\
&& \Big\{(B_{G_{3a}}-B_{G_{3b}}) \Big[-y_{W_H} (y_H+10 y_W-1) - (y_W-1)(y_H+y_W)-y_{W_H}^2\Big] \nonumber \\
&& -\, C_{G_{3a}} (y_H-1) y_W \Big[y_H (1-y_W-5y_{W_H}) + y_W^2+10 y_W y_{W_H}+y_W+y_{W_H}^2+5 y_{W_H}-2\Big] \nonumber \\
&& -\, C_{G_{3b}} (y_H-1)y_{W_H} \Big[y_H (1-5 y_W-y_{W_H})+y_W^2+5 y_W (2y_{W_H}+1)+y_{W_H}^2+y_{W_H}-2\Big]\nonumber\\
&& -\, (y_H-1) \left(y_{W_H} (y_H+10 y_W-1)+(y_W-1)(y_H+y_W)+y_{W_H}^2\right) \Big\},
\nonumber \\
\mathcal{B}^{LO}_{G_3} &=& \frac{2}{(y_H-1)} \mathcal{A}_{G_{3}}, \nonumber \\
\mathcal{C}^{LO}_{G_{3}} &=& 0,
\label{intGb}
\end{eqnarray}
where $C_{G_3}= \dfrac{1}{2cs} \left[ g^4 s_W v (c^2-s^2) \right]$, and
\begin{eqnarray}
B_{G_{3a}} &=& B_0(m_H^2,m_W^2,m_{W_H}^2), \nonumber \\
B_{G_{3b}} &=& B_0(m_{Z_H}^2,m_W^2,m_{W_H}^2), \nonumber \\
C_{G_{3a}} &=& m_{Z_H}^2 C_0(m_H^2,m_{Z_H}^2,0,m_W^2,m_{W_H}^2,m_W^2), \nonumber \\
C_{G_{3b}} &=& m_{Z_H}^2 C_0(m_H^2,m_{Z_H}^2,0,m_{W_H}^2,m_W^2,m_{W_H}^2).
\end{eqnarray}

\par
The $\mathcal{A}^{LO}_{S_{1}}$ and $\mathcal{B}^{LO}_{S_{1}}$ coefficients for loop diagrams contributed by scalars and scalars plus gauge bosons are presented as
\begin{eqnarray}
\label{eq-4}
\mathcal{A}^{LO}_{S_{1}}
&=&
C_{S_{1}} \frac{1}{16\pi^2(y_H-1) y_W}
\Big[ (B_{S_{1a}}-B_{S_{1b}}) (y_H+y_W-y_\phi) +\, C_{S_{1a}} (y_H-1) y_\phi (y_H+ y_W-y_\phi)\nonumber\\
&& -\, C_{S_{1b}} (y_H-1) y_W (y_H+y_\phi-y_W-2) + (y_H-1)(y_H+y_W-y_\phi) \Big], \nonumber\\
\mathcal{B}^{LO}_{S_{1}} &=&\frac{2}{(y_H-1)}\,\mathcal{A}_{S_{1}}, \nonumber\\
\mathcal{C}^{LO}_{S_{1}} &=& 0,
\end{eqnarray}
where $C_{S_1} = \dfrac{1}{2cs} \left[ e g^3 \left(c^2 - s^2\right) v^{\prime}(\sqrt{2}s_0-s_p) \right]$, $y_\phi=m_\phi^2/m_{Z_H}^2$ and
\begin{eqnarray}
B_{S_{1a}} &=& B_0(m_H^2,m_W^2,m_{\phi}^2), \nonumber \\
B_{S_{1b}} &=& B_0(m_{Z_H}^2,m_W^2,m_{\phi}^2), \nonumber \\
C_{S_{1a}} &=& m_{Z_H}^2 C_0(m_H^2,m_{Z_H}^2,0,m_{\phi}^2,m_W^2,m_{\phi}^2), \nonumber \\
C_{S_{1b}} &=& m_{Z_H}^2 C_0(m_H^2,m_{Z_H}^2,0,m_W^2,m_{\phi}^2,m_W^2).
\end{eqnarray}
The mixing angle $s_p$ in the pseudoscalar and singly-charged sectors can be easily extracted in terms of the VEVs, $s_p = \dfrac{2 \sqrt{2} v^{\prime}}{\sqrt{v^2 + 8 v^{\prime 2}}}	\simeq 2 \sqrt{2} \dfrac{v^{\prime}}{v}$. The $\mathcal{A}^{LO}_{S_{2}}$ and $\mathcal{B}^{LO}_{S_{2}}$ coefficients can be obtained by doing the replacement of $m_W \to m_{W_H}$ and $C_{S_1}\to C_{S_2}$ in Eq.(\ref{eq-4}), where $C_{S_2} = \dfrac{1}{4 c^3 s^3} \left[ e g^3 (c^2 - s^2) \left(c^4 + s^4\right) v^{\prime} (\sqrt{2}s_0-s_p) \right]$.

\end{document}